\documentstyle[aps,prl,multicol,epsf]{revtex}

\begin{document}
\title{Disorder Induced Ferromagnetism in Restricted Geometries}

\author
{E. Eisenberg and R. Berkovits}

\address{
The Minerva Center for the Physics of Mesoscopics, Fractals and Neural 
Networks,\\ Department of Physics, Bar-Ilan University,
Ramat-Gan 52900, Israel}

\date{\today}
\maketitle

\begin{abstract}
We study the influence of on-site disorder on the magnetic properties
of the ground state of the infinite $U$ Hubbard model for restricted
geometries. We find that for
two dimensional systems disorder enhances the spin polarization
of the system . The tendency of disorder to enhance magnetism in the
ground state may be relevant to recent experimental observations of
spin polarized ground states in quantum dots and small metallic grains.
\end{abstract}


\begin{multicols}{2}
\narrowtext

The interplay between disorder and interactions\cite{review} 
and the possibility that it leads to ground state ferromagnetism
has been the subject of much interest\cite{para}.
In several new experiments in restricted geometries, such as zero temperature
transport measurements of the conductance through semiconducting 
quantum dots\cite{patel} and carbon nanotubes\cite{tans}, tantalizing
hints of a weakly ferromagnetic ground state of small systems with 
a few hundreds 
of electrons have appeared. 
The ground state spin polarization may be directly measured by
coupling the dot or tube to external leads and  measuring the differential
conductance. Also from a recent mean field treatment of 
electron-electron interactions
in disordered electronic systems\cite{andreev} 
as well as from a numerical study of such systems\cite{ber} 
a partially magnetized ground state seems probable.

The canonical model for the study of itinerant ferromagnetism
is the Hubbard model, described by the Hamiltonian
\begin{eqnarray}
H=\sum_{i\sigma} \varepsilon_i n_{i\sigma}
-t\sum_{<ij>\sigma}a_{i\sigma}^\dagger a_{j\sigma}
+  c.c. +U\sum_i n_{i\uparrow}n_{i\downarrow},
\label{hamil}
\end{eqnarray}
where $a_{i\sigma}^\dagger$ is the fermionic creation operator on site
$i$ with spin $\sigma$, $n_{i\uparrow}=a_{i\uparrow}^\dagger
a_{i\uparrow}$, and the on-site energies $\varepsilon_i$ are drawn
randomly according to a uniform distribution between $-W/2$ and $W/2$.
The large $U$ regime of the model has attracted much interest 
due to its relevance to the theory of high-$T_c$ superconductivity
\cite{anderson}. 
Although the model clearly does not contain many of
the physical attributes of the typical experimental system such as a quantum
dot (especially at the infinite $U$ limit), nevertheless, 
it is important to gain 
insight into the complicated problem of the influence of disorder on
the spin structure of interacting electrons in restricted geometries
by studying simplified models. Moreover, the infinite $U$ limit has
the attractive feature of suppressing antiferromagnetic correlations
which are clearly not relevant to quantum dots, even in the clean limit
\cite{egger}. 

%Among the various types of magnetic order which were considered 
%for the clean Hubbard model \cite{dagotto}, the most simple one is 
%the possibility of itinerant ferromagnetism. We thus choose to study 
%the influence of disorder on the GS magnetic moment. 

In this paper we study the effect of disorder on the enhancement of
the spin polarizability of small interacting systems. 
For one dimensional infinite $U$ Hubbard models, 
disorder does not essentially change the magnetization behavior \cite{eb}. 
On the other hand, for two dimensional systems, it is found that 
sufficiently strong disorder suppresses the singlet-favoring effects
and the spin polarization of the system increases. 
From analytical arguments and numerical calculations it is shown 
that in the presence of disorder the fully polarized regime extends 
to higher densities of holes. Even beyond the fully polarized regime
disorder creates a tendency towards non-zero magnetic moments in
the ground state, which is consistent with other indications for such
a behavior in disordered interacting systems.

The extensively studied Hubbard model, which is the simplest model
of strongly correlated electrons, was originally introduced
to explain ferromagnetism \cite{hubbard}. However, till now, little is
known about the phase diagram of the model even at $T=0$. There are a 
few rigorous results, mostly restricted to the one-dimensional model,
or to the half-filled case. Lieb and Mattis have proven that 
for one dimensional systems (1D) for even number of electrons, 
with interaction strength $U<\infty$ and open boundary conditions
(BC) \cite{lieb} the ground state (GS) is a singlet.
For higher dimensions and half filling 
it has been shown \cite{hf} that for a bipartite lattice with $N_A$ 
($N_B$) sites on sub-lattice A(B), the GS is non-degenerate and has 
total spin ${1\over 2}|N_A-N_B|$, as long as $U>0$. In particular,
for the square lattice, the GS is a singlet. In the large-$U$ limit
the problem is mapped onto the Heisenberg Hamiltonian, leading to
antiferromagnetism (AFM) with long range order \cite{afm}.

An important milestone in the research of ferromagnetism in Hubbard 
models is the work of Nagaoka \cite{nagaoka}. It showed that for most 
lattices, with nearest-neighbor hopping and infinite on-site interaction 
$U$, the GS is the fully saturated ferromagnetic state, for the case
of one hole in an otherwise half-filled band. An extensive work was done
in order to find whether this result can be extended to higher hole density,
or to finite $U$. It was shown that the two-hole case the GS is a singlet 
\cite{doucot,mikh}, but this GS is degenerate with the Nagaoka state
in the thermodynamic limit. Various variational wavefunctions were suggested
to test the stability of the Nagaoka state (See 
\cite{shastry,vdl,wurth,hanisch,okabe} and references therein).
Bounds were given to the
holes density for which stability may remain. The best bound to date is
$\delta_{cr}\leq 0.2514$ \cite{wurth} (where $\delta$ is the number
of holes per site). Still, the stability of the Nagaoka state, 
and the possibility of explaining ferromagnetism by it, is an 
unresolved problem \cite{recent}. 

In this letter we wish to show that in some sense the situation in the 
disordered case is simpler. Let us first sketch the situation in 1D.
If periodic BC are imposed in 1D, the problem of $m$ interacting 
electrons (at $U=\infty$) can be mapped onto a system of
$m$ non-interacting spinless fermions
on 1D ring with fluxes $\Phi/\Phi_0=2\pi j/m$ ($j=0,,m-1$),
where the GS energy corresponds to the flux $j$ with
the lowest energy\cite{eb}.
This mapping shows that the effect of the spin background in 1D is trivial,
and does not depend of the strength of disorder. 

One might expect the effects of disorder in 2D to be smaller than in 1D, and
thus no influence of disorder in 2D as well.
However, the insensitivity of the 1D GS spin structure to disorder
is accounted for by the fact that the spin permutation subgroup induced 
by the 1D hopping terms is cyclic \cite{eb}. Hence, spin background 
effects in 1D are not major. On the other hand, the permutation 
subgroup induced by the 2D hopping terms is
non-abelian, and therefore the spin background has non trivial effects
on the dynamics of the holes. Thus we might expect an interplay between 
disorder and the behavior of the spin background in 2D. 

We start by considering the influence of disorder on the one hole case. 
The hopping of the hole around the lattice induces permutations in the spin
ordering. The hopping term is then effectively reduced by a factor
proportional to the expectation value of the different permutations.
In order to minimize kinetic energy, this overlap should be maximal, and
this is achieved by the fully polarized state for which the spin 
wavefunction is unchanged by permutations of different spins.
This argument does not change due to disorder, and
it leads to the Nagaoka's theorem which assures us that the GS 
is fully polarized, even in the presence of disorder. 

The two holes case is much more complicated. Although the above argument 
for preferring the FM order applies for the case of several 
holes equally, it is known that
in the ordered case the GS is a singlet \cite{doucot,mikh}. 
This is accounted for 
by the re-ordering of the spin background in order to mask the fermionic BC
between the holes. Thus, the spin background takes care for the 
anti-symmetrization of the many body (in fact, many-holes) wave function, 
and thus the spatial function has less nodes, which 
decrease its energy. It was shown, for a special variational wave function, 
that the resulting energy gain supersedes the energy increase at the bottom
of the band, coming from the reduction of the hopping 
amplitude due to the Nagaoka effect \cite{doucot}. Thus, the tendency
towards ferromagnetism is suppressed, and the spin structure, if any,
is of a much more complex form \cite{dagotto}.

As disorder increases, the single particle functions  become more and more
localized (in the participation ratio sense). The overlap of the different
single particle functions decreases, and thus, 
the fermionic BC constraint (i.e., a zero of the many body wavefunction 
wherever two particles are on the same site) becomes less restrictive, and 
does not change the many body energy much. 
Therefore, one may expect that the incentive for re-ordering of the spin 
background decreases, while, as in the one hole case, there still is
a contribution from the hopping amplitude leading to a Nagaoka state.

\begin{figure}
\centerline{\epsfxsize = 1.7in \epsffile{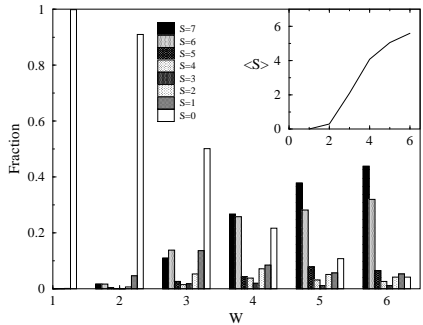}}
\caption{The spin distributions as a function of disorder $W$
for a $4\times4$ lattice with 14 electrons. 
For each $W$, the bar chart represents the probability of finding the
GS of the system at a particular value of $S$.
The inset presents the average spin $\langle S \rangle$ as a function of 
$W$.}
\label{f4x4}
\end{figure}

Exact diagonalization for the full many-particle Hamiltonian of 
Eq. (\ref{hamil}) was used to test the above arguments. 
Although we have used small systems one may expect that due to 
the chaotic nature of the dots \cite {ber} the dependence on the 
number of electrons or the BC will play a less important role
for disordered systems than in clean ones \cite{dagotto}. Thus,
the study of a small number of electrons is still useful in understanding
the properties of dots which are populated by an order of magnitude more 
electrons. We have used
up to $14$ electrons on up to $4\times 4$ lattices. The size of
the Hilbert space is then $471435600$, which is far beyond
exact diagonalization capabilities. Fortunately one can omit the 
double occupied states for $U=\infty$ and use the spin symmetry of the 
Hamiltonian to reduce this number considerably.
The number of spatial functions in this case is 
$120$, and the number of total spin configurations in the $S_z =0$ sector
is $3432$, yielding a total of $411840$ states. We have used group theory to
construct the definite $S$ states, and to decompose the space into subspaces
of definite $S$ and $S_z$. The largest sectors ($S=1,2$) consisted of $1001$
spin functions and a total of $120120$ basis functions. Group theory was
used for constructing the matrices describing the effect of hopping on
the different spin functions. We then employed the Lanczos algorithm to find
the exact GS for $600$ realizations at every disorder value.
In the ordered case, the GS was a singlet,
in accordance with \cite{doucot,mikh}. 
Figure \ref{f4x4} presents the GS-spin distributions as a function of $W$, 
for $14$ electrons on a hard-wall $4\times 4$ lattice. The average spin 
$\langle S \rangle$ is also plotted
against $W$, and one can see that it increases significantly with $W$.
In the presence of disorder, one gets 
a distribution of GS-spin values. For weak disorder, the main effect is
smearing the peak at $S=0$ to low $S$ values. Thus, a tendency
towards weak ferromagnetism is clearly demonstrated even for weak
disorder ($W=3t$) which corresponds to a ballistic (mean free path larger
than the system size) regime. Moreover, as disorder
increases, high $S$ values dominate the distribution.
For $W=6t$ corresponding to a diffusive regime a clear dominance of
the high spin state appears.

\begin{figure}
\centerline{\epsfxsize = 1.7in \epsffile{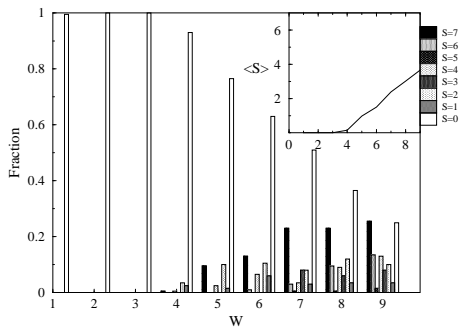}}
\caption{Same as Fig. \protect\ref{f4x4} for periodic BC.}
\label{f4x4p}
\end{figure}

Similar behavior was obtained for smaller lattices and periodic BC.
Fig. \ref{f4x4p} presents the results for the same conditions as
in Fig. \ref{f4x4} employing periodic BC. Clearly,
the tendency towards ferromagnetic behavior persists, although
higher values of $W$ needed to obtain similar values of spin polarization. 
This is the result of the fact that for
periodic BC, higher values of $W$ are needed to 
generate the same value of dimensionless conductance.
One sees that, in contrast with the situation in the ordered case,
our results are not sensitive to the lattice size or the BC.
This manifests the chaotic nature of the dot, which suppress
dependencies on the details of the system.

A clear manifestation of this point is presented by
the results for $13$ electrons on
a $5\times 3$ lattice. In the ordered case, the behavior of this
cluster depends dramatically on the BC. For hard-wall BC, the GS 
is fully polarized (i.e., $S=13/2$), while for periodic BC, the GS has
the minimal spin $S=1/2$. On the other hand, once the system is diffusive 
the GS-spin polarization distributions become closer and when 
the dimensionless 
conductance is of order one, both  distributions are quite similar, where 
$\langle S \rangle = 5.90$
for hard-wall BC and  $\langle S \rangle = 3.96$ for periodic BC.

\begin{figure}
\centerline{\epsfxsize = 1.7in \epsffile{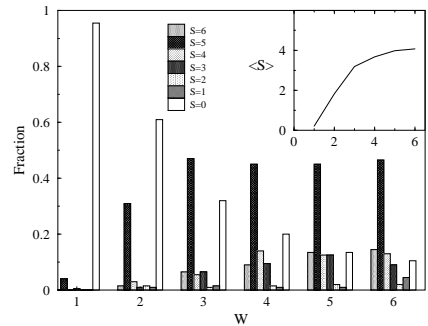}}
\caption{The spin distributions as a function of disorder $W$
for a $5\times3$ lattice with 12 electrons.}
\label{f5x3}
\end{figure}

Exact diagonalization also confirms the tendency towards non-zero
ground state spin values even for a higher number of holes. In Fig.
\ref{f5x3} we depict the spin distribution for $12$  electrons
on a hard-wall $5\times 3$ lattice (3 holes). The GS-spin is significantly
enhanced as function of disorder, although the most probable spin state is
not fully ferromagnetic. This tendency towards partial polarization of the
ground state persists in higher hole ratios.

\begin{figure}
\centerline{\epsfxsize = 1.7in \epsffile{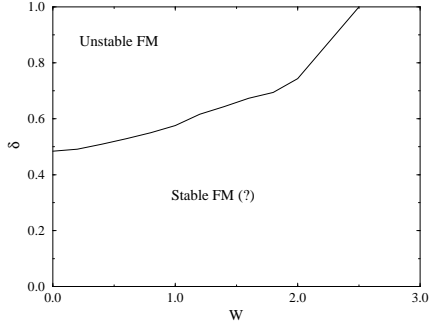}}
% figure on /nagaoka on neuron using file finalmap
% taken from albert /nagaoka/newcomp/better
\caption{Stability curve for the single flip excitation: the critical
hole density $\delta$ vs. the disorder distribution width $W$} 
\label{map}
\end{figure}

The method of exact diagonalization is restricted to small lattices. 
In order to learn whether the tendency towards ferromagnetism persists for
larger systems we turn to a variational method.
Many authors have considered various variational wave functions to study the 
instability of the Nagaoka state of the $U=\infty$ model for a thermodynamic 
concentration of holes \cite{shastry,vdl,wurth,hanisch,okabe}. 
Since the reliability of these functions for an accurate calculation
of the phase boundary of ferromagnetism is doubtful, we only use
this method to get a hint about disorder influence of the stability.
For this purpose, we use
the most simple of these functions\cite{shastry}, which is one of a 
single particle excitation. An up spin electron is removed from the occupied
states and placed with flipped spin into another state. 
Direct calculation of the excitation energy in the ordered case
\cite{shastry} yields stability of the Nagaoka state with respect to spin flip
for $\delta=0.49$ (for a square lattice), while for smaller hole concentration, 
the Nagaoka state remains stable with respect to this excitation. 
We have done the calculation for the disordered case by
taking different realizations of disorder of a $24\times 24$ system, 
diagonalizing the single particle (non-interacting) Hamiltonian to find its 
eigenvalues and eigenvectors, and then calculating directly the excitation 
energy of the single flip variational wave function. Figure \ref{map}
shows the stability regime in the $\delta-W$ plane, 
as follows from this 
excitation calculation. For an ordered system we get the result of Ref. 
\cite{shastry} that the ferromagnetic state is stable for $\delta\leq 0.49$. 
However, as disorder increases, the stability regime grows.
It therefore seems that the exact results for small systems characterize 
the behavior in larger systems as well. We note that analytical 
calculation of the excitation energy in the disordered case
according to random vector model (RVM) yields a completely different
behavior. This stems from the fact that RVM ignores correlations
between the wavefunctions and the eigenvalues which are important in
this case.

We would like to add a remark about the $U=\infty$ limit. The Nagaoka's 
effect and the decrease in the singlet favoring effect described here, 
are not unique to the $U=\infty$ limit. However, for the Hubbard model, 
ferromagnetism arises (even for one hole) only for $U\gg t$. The reason is 
that due to the perfect nesting property of the lattice model, the
GS of the almost half filled case tends to be AFM. In order to wash
out this tendency, the limit $U=\infty$ is taken.
However, in reality,  quantum dots do not show AFM behavior, 
since they are not described by a perfect lattice.
One then might expect the previously described effect, namely the 
formation of larger magnetic moments due to disorder, to show in
real quantum dots even for moderate values of $U$.

In conclusion, the influence of disorder on the magnetic properties
of the GS was studied. For an ordered system, large magnetic moments 
are generally suppressed, and the spin structure of the GS, if any, 
is very complicated. On the other hand, we have shown that disorder 
plays an important role in determining the spin polarization of 2D 
systems described by the infinite $U$ Hubbard model. Weak disorder 
tends to create a partially polarized ground state, while stronger 
disorder tends to stabilize a fully ferromagnetic GS. This behavior 
clearly indicates that there is a basis to expect that for more 
realistic descriptions of the experimental systems ($U\ne\infty$) 
disorder will play an important role in creating a spin polarized 
ground state.

We would like to thank The Israel Science
Foundation Centers of
Excellence Program and the Clore Foundation
for financial support.

\end{multicols}
\end{document}